\RequirePackage{fix-cm}
\RequirePackage{amsmath}
\documentclass[twocolumn,epjc3]{svjour3}  

\smartqed  
\RequirePackage{graphicx}
\RequirePackage{mathptmx}      
 
\usepackage{color}
\usepackage{mathrsfs}
\usepackage{amssymb, bm}

\newcommand{\be}{\begin{equation}}
\newcommand{\ee}{\end{equation}}
\def\Box{\hbox{$\rlap{$\sqcup$}\sqcap$}}
\begin{document}

\title{More on the first-order thermodynamics of scalar-tensor 
and Horndeski gravity 
}


\author{Valerio Faraoni\thanksref{e1,addr1}
	   \and
Julien Houle \thanksref{e2,addr1} 
}

\thankstext{e1}{e-mail: vfaraoni@ubishops.ca}
\thankstext{e2}{e-mail: jhoule22@ubishops.ca}


\institute{Department of Physics \& Astronomy, Bishop's University, 
2600 College Street, Sherbrooke, Qu\'ebec, Canada J1M~1Z7 \label{addr1}
}

\date{Received: date / Accepted: date}

\maketitle

\begin{abstract}

Two issues in the first-order thermodynamics of scalar-tensor (including 
``viable'' Horndeski) gravity are elucidated. The application of this new 
formalism to FLRW cosmology is shown to be fully legitimate and then 
extended to all Bianchi universes. It is shown that the formalism holds 
thanks to the almost miraculous fact that the constitutive relations of 
Eckart's thermodynamics are satisfied, while writing the field equations 
as effective Einstein equations with an effective dissipative fluid does 
not contain new physics.

\end{abstract}

\section{Introduction}
\label{sec:1}
\setcounter{equation}{0}

There are many motivations to consider seriously theories of gravity 
alternative to General Relativity (GR) \cite{Faraoni:2010pgm}. 
Attempts to quantum-correct GR 
generally lead to new degrees of freedom (in addition to the two familiar 
massless spin two modes), higher order equations of motion, extra 
fields, or non-local features. The low-energy limit of the bosonic string, 
the simplest string 
theory, does not reproduce Einstein gravity but gives an $\omega=-1$ 
Brans-Dicke theory instead (where $\omega$ is the Brans-Dicke parameter) 
\cite{Callan:1985ia,Fradkin:1985ys}. More 
compelling motivation comes from the accelerated expansion of the 
universe: explaining the present-day cosmic acceleration within 
the context of GR requires the introduction of a completely {\em ad hoc} 
dark energy, the nature of which remains mysterious 
\cite{AmendolaTsujikawabook}.

The simplest alternative to GR is scalar-tensor gravity, which adds only a 
scalar degree of freedom $\phi$ to the two degrees of freedom contained in 
the metric 
$g_{ab}$ in GR. A subclass of 
scalar-tensor gravity, $f(R)$ theories, seems to be the most popular 
alternative to GR to explain the cosmic acceleration 
(\cite{Capozziello:2002rd,Carroll:2003wy}, see 
\cite{Sotiriou:2008rp,DeFelice:2010aj,Nojiri:2010wj} for reviews). During 
the past decade Horndeski gravity \cite{Horndeski:1974wa} 
was rediscovered while attempting to write down the most general 
scalar-tensor theory with second-order equations of motion. Although 
this record ultimately belongs to the newly-discovered Degenerate Higher 
Order 
Scalar-Tensor (DHOST) theories more general than 
Horndeski's, the study of Horndeski gravity has flourished, generating a 
large literature (see, {\em e.g.}, 
\cite{H1,H2,H3,GLPV1,GLPV2,DHOST1,DHOST2,DHOST3,DHOST4,DHOST5,DHOST6,DHOST7,DHOSTreview1,DHOSTreview2,Creminellietal18,Langlois:2018dxi,Kobayashi:2011nu,Noller,Afshordi:2014qaa,Kobayashi:2016xpl,Akama:2017jsa,Creminelli:2016zwa,Panpanich:2021lsd,Starobinsky:2019xdp} for 
its various aspects, including cosmology, and for reviews). 

At the same time, the idea that gravity may be different from the other 
three fundamental forces and may be emergent instead, has taken a firm 
foot in the literature in various forms  (see 
\cite{Sakharov,Visser:2002ew,Padmanabhan:2008wi,Padmanabhan:2009vy,Hu:2009jd,Verlinde:2010hp,Carlip:2012wa,Giusti:2019wdx} for reviews). 
A particularly deep approach is Jacobson's thermodynamics 
of spacetime, in which the Einstein equation is derived from  
thermodynamics \cite{Jacobson:1995ab}. When applied to (metric) $f(R)$ 
gravity \cite{Eling:2006aw}, it embodies the idea that GR constitutes a 
state of equilibrium for gravity while $f(R)$ gravity is an out-of 
equilibrium state (\cite{Eling:2006aw}, see also \cite{Chirco:2010sw}). 
This idea has been adopted in the completely different first-order 
thermodynamics of scalar-tensor gravity recently proposed 
\cite{Faraoni:2018qdr,Faraoni:2021lfc,Faraoni:2021jri,Giusti:2021sku,Giardino:2022sdv}. This new formalism begins from the  
realization that the field equations of ``first-generation'' 
scalar-tensor 
\cite{Jordan38,Jordan:1959eg,Brans:1961sx,Bergmann:1968ve,Nordtvedt:1968qs,Wagoner:1970vr,Nordtvedt:1970uv}
 and Horndeski gravity can be rewritten in the form of 
effective Einstein equations (using the notations of 
Ref.~\cite{Wald:1984rg})
\be
R_{ab}-\frac{1}{2} \, g_{ab} R = 8\pi \left( T_{ab}^{(\phi)} + 
\frac{ T_{ab}^\mathrm{(matter)} }{\phi} \right) \,,
\ee
where $R_{ab}$ is the Ricci tensor of the metric $ g_{ab}$ and 
$T_{ab}^\mathrm{(matter)}$ is the matter stress-energy tensor, while 
$T_{ab}^{(\phi)}$ is an effective stress-energy tensor built out of the 
gravitational scalar $\phi$ and its first and second covariant 
derivatives (indeed, this is the usual way to present ``first 
generation'' scalar-tensor gravity). Furthermore, the effective 
$T_{ab}^{(\phi)}$ 
assumes the form of an imperfect fluid energy-momentum 
tensor
\be
T_{ab}= \rho u_a u_b +P h_{ab} +\pi_{ab} + q_a u_b + q_b u_a 
\,,\label{imperfectfluid}
\ee
where $ u^a$ is the fluid 4-velocity (normalized to $u_c u^c=-1$), $h_{ab} 
\equiv g_{ab}+u_a u_b$ is the metric of the 3-space seen by the observers 
with 4-velocity $u^a$ comoving with the fluid, $\rho $ is the energy 
density, $P$ the isotropic pressure, $\pi_{ab}$ is the anisotropic stress 
tensor, and $q^a$ is a heat flux density. $P = \bar{P}+P_\mathrm{viscous}$ 
is the sum of a perfect fluid contribution $\bar{P}$ and of a viscous 
pressure; $h_{ab}, \pi_{ab}$, and $q^a$ 
are purely spatial,
\be
h_{ab} u^a=h_{ab}u^b=q_a u^a =\pi_{ab} u^a =\pi_{ab}u^b=0 \,,
\ee
and $\pi_{ab}$ is trace-free, ${\pi_a}^a=0$. 

The dissipative fluid nature of the effective scalar field stress-energy 
tensor was recognized in \cite{Pimentel89,Faraoni:2018qdr}  for ``old'' 
scalar-tensor gravity and in \cite{Quiros:2019gai,Giusti:2021sku} for 
Horndeski gravity. In fact, {\em 
any} symmetric two-index tensor can be decomposed in the 
form~(\ref{imperfectfluid}) (more on this in Sec.~\ref{sec:4}).  What 
makes the  
analogy between scalar field and dissipative fluid meaningful is the fact 
that the 
effective fluid quantities satisfy the  constitutive relations of Eckart's  
first-order thermodynamics \cite{Eckart:1940te} 
\begin{eqnarray}
&& q_a = - K h_{ab} \left( \nabla^b {\cal T} + {\cal T} \dot{u}^b  
\right) \,,\label{Eckart1} \\
&&\nonumber\\
&& \pi_{ab} = -2\eta \, \sigma_{ab} \,,\label{Eckart2}\\
&&\nonumber\\
&& P_\mathrm{viscous} =  -\zeta \, \Theta \,,\label{Eckart3}
\end{eqnarray}
where ${\cal T}$ is the temperature, $K$ is the thermal 
conductivity, and $\Theta, \sigma_{ab}$ are the expansion and trace-free 
shear of the 4-velocity $u^a$, while $\dot{u} \equiv u^c\nabla_c u^a$ is 
the fluid 4-acceleration. 
The fact that the {\em effective} $T_{ab}^{(\phi)}$ satisfies 
Eqs.~(\ref{Eckart1})-(\ref{Eckart3})  was realized for general ``old'' 
scalar-tensor gravity in 
\cite{Faraoni:2018qdr} (and, for particular geometries or theories, in 
previous references \cite{Madsen:1988ph}) and in ``viable'' Horndeski 
gravity in \cite{Quiros:2019gai,Giusti:2021sku} and identifies a 
``temperature of gravity''  with respect to GR. Einstein gravity, 
recovered for 
$\phi=$~const., corresponds to zero temperature while 
scalar-tensor gravity is an excited state. This idea is plausible: if the 
field content of gravity consists of the two massless spin two modes of GR 
plus a propagating scalar mode, exciting the latter corresponds to an 
excited state with respect to GR. The whole idea of the first-order 
thermodynamics of scalar-tensor gravity consists of taking seriously the 
dissipative form of the effective $T_{ab}^{(\phi)}$ and applying Eckart's  
thermodynamics to it. It is something akin to a  miracle that Eckart's 
constitutive 
relations are satisfied \cite{Faraoni:2018qdr,Faraoni:2021lfc,Giusti:2021sku}. 
With all the limitations intrinsic to Eckart's  thermodynamics (lack of 
causality and instabilities 
\cite{Maartens:1996vi,Andersson:2006nr}), an 
intriguing thermal picture of modified gravity emerges 
\cite{Faraoni:2018qdr,Faraoni:2021lfc,Faraoni:2021jri,Giusti:2021sku,Giardino:2022sdv}, 
which is under development.  Ideas and tools partially or fully 
developed include: an explicit equation describing the approach to the GR 
equilibrium or the departures from it; the expansion of space causes the 
``cooling'' of gravity; near spacetime singularities and singularities of 
the effective gravitational coupling, where the scalar degree of freedom 
is fully excited, gravity is ``hot'' and deviates radically from GR; in 
cosmology only bulk viscosity 
survives 
due to the spacetime symmetries; states of equilibrium corresponding to 
$ K{\cal T}=0$ (or even $ K {\cal T}=$~const.) distinct from GR 
can exist, corresponding to non-dynamical scalar fields or to metastable 
states 
\cite{Faraoni:2022doe,Faraoni:2022jyd,Faraoni:2022fxo}.\footnote{The  
construction of an 
effective  $T_{ab}$ extends 
to Nordstr\"{o}m gravity, which is not a scalar-tensor but a purely scalar 
theory with less degrees of freedom than GR, and yields a negative 
temperature with respect to GR \cite{Faraoni:2022doe}.}

We summarize Horndeski theory for use in the following sections.  Denoting 
$X \equiv -\frac{1}{2} \, \nabla^c \phi \nabla_c \phi$, the 
Lagrangian of Horndeski gravity reads 
\be
\mathcal{L} = \mathcal{L}_2 + \mathcal{L}_3 + \mathcal{L}_4 + 
\mathcal{L}_5 \,,
\ee
where 
\begin{eqnarray}
\label{eq:HorndeskiGeneral}
 \mathcal{L}_2 &=& G_2 \left( \phi, X \right) \, , \\
&&\nonumber\\
\mathcal{L}_3 &=& - G_3\left( \phi, X \right)  \, \Box \phi \, ,\\
&&\nonumber\\
\mathcal{L}_4 &=& G_4\left( \phi, X \right)  \, R 
+ G_{4 X}\left( \phi, X \right)  \left[ (\Box \phi)^2 - 
(\nabla_a \nabla_b \phi)^2 \right] \,,\nonumber\\
&&\\
\mathcal{L}_5 &=& G_5\left( \phi, X \right)   \, G_{ab} \, 
\nabla^a \nabla^b 
\phi -  \frac{G_{5X}\left( \phi, X \right) }{6} \nonumber\\
&&\nonumber\\
&\, &  \times  \Big[  (\Box \phi)^3   - 3 \, \Box \phi \, 
(\nabla_a \nabla_b \phi)^2  + 2 \, (\nabla_a \nabla_b \phi)^3 \Big] 
\,,\nonumber\\
&&
\end{eqnarray}
and where  $\nabla_a $ is the covariant derivative of  $g_{ab}$, $\Box 
\equiv  g^{ab} \nabla_a \nabla_b $, $G_{ab}$ is the 
Einstein tensor, while $G_i (\phi, X)$ ($i=2,3,4,5$) are arbitrary 
functions of $\phi$ and  $X$, while   
$G_{i\phi} \equiv \partial G_i /  \partial  \phi$,  ~$G_{iX} \equiv 
\partial G_i / \partial  X$.

Horndeski gravity is constrained  theoretically by  the 
need to avoid graviton decay into 
scalar field perturbations \cite{Creminellietal18} and, above all, by the 
2017 multi-messenger observation of gravitational waves and 
$\gamma$-rays emitted simultaneously in the GW170817/GRB170817 event 
\cite{TheLIGOScientific:2017qsa,Monitor:2017mdv}, which sets a stringent  
upper limit on the difference between the propagation speeds 
of gravitational and electromagnetic waves \cite{Langloisetal18}. The 
subclass of Horndeski theories that 
implies luminal propagation of gravitational waves has $G_{4 X} = 0$, 
$G_5=0$ and its Lagrangian density is  restricted to
\be \label{eq:LsH}
\mathcal{\bar{L}}= G_2 (\phi, X) - G_3 (\phi, X) \Box \phi + G_4 (\phi) R 
\,. 
\ee

\section{Correct generalization of Fourier's law}
\label{sec:2}

Eckart's generalization of Fourier's law \cite{Eckart:1940te} is often 
reported as 
\be 
q_a =- K \left( h_{ab} \nabla^b {\cal T} + {\cal T} 
\dot{u}_a \right) \,.\label{questa} 
\ee 
The heat flux density $q^a$ is 
purely spatial in Eckart's theory and, therefore, non-causal, an 
unphysical feature corrected in the Israel-Stewart second-order 
thermodynamics and in later formalisms. While, in the right-hand side of 
Eq.~(\ref{questa}), $- K h_{ab} \nabla^b {\cal T}$ is trivially a 
purely spatial vector (it is a projection onto the 3-space orthogonal to 
$u^a$), the second term $-  K {\cal T} \dot{u}^a$ proportional to 
the fluid 4-acceleration is not always a spatial vector, contrary to 
intuition. While most of the times a particle's 4-acceleration is 
orthogonal to the particle 4-velocity, this is not always the case. 
Although at first sight this may seem hair-splitting, relevant situations 
discussed in the literature span a range of interesting subjects including 
particles with varying mass, the Einstein frame of scalar-tensor gravity, 
cosmic antifriction due to self-interacting dark matter or to particle 
production, and Friedmann-Lema\^itre-Robertson-Walker (FLRW)  cosmology 
sourced by a perfect fluid with pressure in the comoving frame 
\cite{Faraoni:2020ejh}. There is an abundant literature on analytic 
solutions of the Einstein equations describing mass-varying sustems such 
as rockets and solar sails in GR ({\em e.g.}, 
\cite{Forward84,Fuzfa:2019djg,Fuzfa:2020dgw} and references therein) and 
mass-changing particles in cosmology and in scalar-tensor gravity 
\cite{Mbelek:1998vu,Mbelek:2004ff,Damour:1990tw,Casas:1991ky,Garcia-Bellido:1992xlz,AndersonCarroll97}. In the early universe, quantum 
processes can create particles, a phenomenon associated with negative bulk 
pressures \cite{Zeldovich:1970si,Hu}, and it has been suggested that such 
a mechanism could drive inflation 
\cite{Zimdahl:2000zm,Schwarz:2001cf,Zimdahl:1996cg,Zimdahl:1997qe,ZimdahlBalakin98a,Zimdahl:1998zq}.  Negative bulk stresses can be caused 
by the self-interaction of dark matter, which has been investigated as a 
possible cause of the present cosmic acceleration \cite{Zimdahl:2000zm} 
because it causes a cosmic ``antifriction'' on the dark matter fluid, a 
force antiparallel to the worldlines of dark matter particles 
\cite{Zimdahl:2000zm}.

In the Einstein conformal frame of scalar-tensor cosmology, a similar 
4-force 
parallel to the trajectory appears. It can be interpreted as due to the 
fact that what was the constant mass of a test 
particle in the Jordan frame now depends on the Brans-Dicke-like scalar 
$\phi$ (that is, upon transformation to the Einstein conformal frame  
massive test particles cease being test particles and are subject to a 
fifth force proportional to $\nabla^a\phi$)  
\cite{Faraoni:2004pi,Faraoni:2020ejh}. In 
the low-energy limit of string theories, the geodesic equation of dilaton 
gravity contains a similar correction but, in general, the coupling of the 
dilaton to particles of the Standard Model is not universal 
\cite{Taylor:1988nw,Damour:1994zq,Gasperini:1999ne}.

Consider a FLRW universe with line element
\be
ds^2 =-dt^2 +a^2(t) \left[ \frac{dr^2}{1-kr^2} +r^2 \left( d\vartheta^2 
+\sin^2\vartheta \, d\varphi^2 \right)\right]
\ee
in comoving coordinates $\left( t, r, \vartheta, \varphi \right)$, that 
is, in the frame 
adapted to the symmetries (spatial homogeneity and isotropy) and comoving 
with 
the perfect fluid usually causing gravity. Unless this fluid is a dust 
or the effective fluid $T_{ab}^{(\Lambda)}=-\frac{\Lambda}{8\pi} 
\, g_{ab}$ describing a cosmological constant term with 
constant pressure, there are a pressure $P(t)$ and a pressure 
gradient $\nabla_a P \neq 0$, hence a 4-force pointing in the 
time direction $u^a$. The 
presence of this force makes fluid particles deviate from geodesics, hence 
they have a 4-acceleration. This is easy to understand since, due to the 
symmetries, this 4-acceleration and 4-force cannot have spatial components 
in the comoving frame. As a result, the (massive) fluid particles satisfy 
the equation of motion \cite{Faraoni:2020ejh}
\be
\frac{d^2 x^{\mu}}{dt^2} + \Gamma^{\mu}_{\alpha\beta} \, 
\frac{dx^{\alpha}}{dt} \, \frac{ dx^{\beta}}{dt}  
= A \, \frac{dx^{\mu}}{dt} \,, \label{non-affinegeodesic}
\ee
where $A$ is a function of the position on the timelike trajectory. 
Equation~(\ref{non-affinegeodesic}) is recognized as the non-affinely 
parameterized timelike geodesic 
equation with the comoving time $t$ coinciding with the the proper time 
of comoving 
observers. It is always possible to change parametrization to an affine 
parameter in which the right-hand side of the geodesic equation vanishes, 
hence this 4-acceleration is regarded as trivial and usually described as 
vanishing, but the reparametrization entails the use of an affine 
parameter that is not the  
comoving time $t$, which is the proper time of comoving observers. (If $s$ 
is an affine parameter, the function $A$ in 
Eq.~(\ref{non-affinegeodesic}) is $A(t)=  \frac{dt}{ds} \,  
\frac{d^2s}{dt^2} $ \cite{EMMacC,Faraoni:2020ejh}, see~\ref{appendix:A}.) 
In other words, the equation describing 
the spacetime trajectory of the fluid particles cannot be affinely 
parametrized by the proper time of comoving observers and there is a 
4-force parallel to the 4-velocity $u^a$ in this frame 
\cite{Faraoni:2020ejh} (see~\ref{appendix:A}).\footnote{Whether 
this 4-force parallel to the 
trajectory can legitimately be called a ``force'' is a matter of 
semantics. Similarly, the 4-acceleration $\dot{u}^a$ of a particle is 
often taken to be synonimous of its spatial projection $h_{ab} \dot{u}^b$. 
For clarity, here we make the distinction explicit.} While this is 
immaterial for the mathematics of curves, the difference between proper 
time of comoving observers and another parameter is important for the 
physics because the description of FLRW cosmology is always done with 
respect to comoving observers who see the cosmic microwave background 
homogeneous and isotropic around them (apart from tiny temperature 
fluctuations).

It is clear then that, in Eckart's first constitutive 
relation~(\ref{questa}), the term $- K{\cal T} \dot{u}_a$  
contributing to the heat flux density $q_a$ is not always purely spatial 
and, as a result, $q_a$ is not purely spatial, either. This fact is 
important because the effective first-order thermodynamics of 
scalar-tensor and Horndeski gravity \`a la Eckart,  including  
FLRW cosmologies, is based on Eckart's generalization of Fourier's law. It 
is easy to fix Eq.~(\ref{questa}) to make the heat flux density $q_a$ 
purely 
spatial in all situations: it is sufficient to 
write\footnote{Interestingly, 
this correct form appears in Eckart's original discussion 
\cite{Eckart:1940te} in which, however, there is no mention of the 
possibility of 4-accelerations parallel to  4-velocities. Indeed, 
Eckart's work \cite{Eckart:1940te} predates all the literature on 
such instances 
\cite{Brans:1961sx,Forward84,Damour:1990tw,Casas:1991ky,Garcia-Bellido:1992xlz,AndersonCarroll97,Zimdahl:1996cg,Zimdahl:1997qe,ZimdahlBalakin98a,Zimdahl:1998zq,Mbelek:1998vu,Mbelek:2004ff,Zimdahl:2000zm,Schwarz:2001cf,Faraoni:2004pi,Fuzfa:2019djg,Fuzfa:2020dgw,Faraoni:2020ejh}.} 
\be
q_a = - K h_{ab} \left( \nabla^b {\cal T}+{\cal T} \dot{u}^a 
\right) 
\,,\label{Eckartcorrected}
\ee
{\em i.e.}, projecting both temperature gradient $\nabla^b {\cal T}$ and 
the 4-acceleration  of the dissipative fluid onto the 3-space orthogonal 
to $u^a$.

An apparent puzzle remains in the study of the first-order thermodynamics 
of scalar-tensor and Horndeski gravity in FLRW cosmology, which is 
addressed in the next section.

\section{Scalar-tensor thermodynamics in FLRW and in Bianchi cosmology}
\label{sec:3}

The study of Eckart's thermodynamics in FLRW cosmology has been carried 
out for first-generation scalar-tensor gravity \cite{Giardino:2022sdv} and 
is being generalized to spatially anisotropic Bianchi cosmologies and to 
Horndeski gravity. To put these studies on a firm footing, we elucidate 
the validity of its formulas involving the effective 
temperature in cosmology, where the heat flux $q^a$ vanishes identically.

Let $\phi$ be the gravitational scalar degree of freedom of the theory and 
$X \equiv -\frac{1}{2} \, \nabla^c\phi \nabla_c \phi$.  In general, 
scalar-tensor thermodynamics is studied in the comoving 
frame, {\em i.e.}, the frame moving with the effective fluid 4-velocity, 
in which the effective fluid is at rest (this is natural in  
tensor-single-scalar gravity; the analogue of the comoving frame 
becomes artificial in tensor-multi-scalar gravity \cite{Miranda:2022uyk}). 
Applying this formalism to FLRW cosmology, it is clear that the purely 
spatial 
heat flux density~(\ref{Eckartcorrected}) must vanish 
in the comoving frame to respect the FLRW symmetries. However, the 
fundamental relation of this thermodynamics
\be
 K{\cal T} = \frac{\sqrt{2X}}{8\pi \phi} \label{KTST}
\ee
 in ``first 
generation'' scalar-tensor gravity, or its counterpart
\be
K{\cal T} = \frac{\sqrt{2X} \, \left( G_{4\phi} -XG_{3X} 
\right)}{G_4} \label{KTHorndeski}
\ee
in viable Horndeski gravity, are derived from identifying the 
effective heat flux density $q_a$ of Eq.~(\ref{Eckartcorrected})  in these 
theories with $ - K {\cal T} h_{ab} \dot{u}^b$. The relation 
(\ref{KTST}) or (\ref{KTHorndeski}) derived in the general theory still 
holds in FLRW cosmology. $q_a$ vanishes identically in the comoving 
frame not because $ K{\cal T}=0$ but {\em because $h_{ab}\dot{u}^b 
=0$ in any FLRW universe}.

Let us be more specific: in Horndeski gravity, the 4-velocity of the 
effective fluid is
\be
u^a = \frac{\nabla^a \phi}{\sqrt{2X}}
\ee
(the analogy is meaningful if  $\nabla^a\phi$ is timelike and 
future-oriented \cite{Giusti:2022tgq}) and the fluid's 4-acceleration 
turns out to be
\be
\dot{u}^a \equiv u^c \nabla_c u^a = -\frac{1}{2X} \left( \nabla^a X + 
\frac{ \nabla^c X \nabla_c \phi}{2X} \, \nabla^a \phi \right) \,.
\ee
Its projection onto the 3-space orthogonal to $u^a $ vanishes if and only 
if $\dot{u}^a = \alpha \, u^a$ (where $\alpha$ is a function of the 
spacetime coordinates), or
\be
\nabla^a X + \frac{ \dot{X}}{\sqrt{2X}} \, \nabla^a \phi =-2\alpha X u^a 
\ee
(where $\dot{X} \equiv u^c \nabla_c X$), or 
\be
\nabla^a X =-\left( \dot{X} + 2\alpha X \right) u^a \,,\label{4-velocity}
\ee
{\em i.e.}, if $\nabla^a X$ is parallel to the effective fluid 4-velocity. 
In 
a FLRW universe, or in any space in which $g^{00}$ depends only on 
the comoving time $t$ and $\phi=\phi(t)$ we have, in comoving coordinates,  
\be
\nabla_a X = \partial_a X = - \frac{\dot{\phi}}{2} \left( \dot{\phi} 
\, \partial_t g^{00}   + 2g^{00}  \ddot{\phi} \right) {\delta_a}^0 
\,.\label{quella}
\ee
Using $u_a=\frac{\nabla_a \phi}{\sqrt{2X}} = 
\frac{\dot{\phi}}{\sqrt{2X}} \, {\delta_a}^0 $, one writes 
\be
\nabla_a X = - \sqrt{ \frac{X}{2}}  \, 
\left( \dot{\phi} \, \partial_t g^{00} 
+ 2 g^{00} \, \ddot{\phi} \right) u_a \,.
\ee
In the FLRW geometry written in comoving coordinates it is $g^{00}=-1$ and 
$\nabla_a X$ reduces to $ - |\dot{\phi}| \ddot{\phi}  \, u_a $, 
which is indeed 
parallel to $u^a$ and then $h_{ab} \, \dot{u}^b =0$. The heat flux density 
of viable Horndeski gravity \cite{Giusti:2021sku}
\be
q_a= \sqrt{2X} \, \frac{ \left( G_{4\phi}-XG_{3X} \right)}{G_4} \, 
h_{ab} \dot{u}^b =- K{\cal T} h_{ab} \dot{u}^b
\ee
vanishes not because $ K{\cal T}=0$ but because $h_{ab} \dot{u}^b=0$ 
(even though $\dot{u}^b$ is non-vanishing in FLRW cosmology).  

The same situation occurs in Bianchi universes in which, again,  
$g^{00}=-1$ in comoving coordinates. Consider first vacuum Horndeski 
gravity, in which the $\phi$-fluid is the only source in the 
effective Einstein equations. The 4-velocity of this effective fluid  
comes from a gradient, therefore it has zero 
vorticity, $\omega_{ab}=0$. Then the Frobenius theorem guarantees that 
$u^a$ is hypersurface-orthogonal and excludes the possibility that the 
$\phi$-fluid is tilted with respect to the Bianchi observers (the 
observers that see the 3-space of a Bianchi universe as homogeneous) 
\cite{Wald:1984rg,EMMacC}.  
There is a time $t$ (``comoving 
time'') such that the 3-surfaces $t=$~const. are spacelike surfaces of 
homogeneity, $\phi=\phi(t)$, and $u^a$ coincides with the unit normal to 
these hypersurfaces. Denoting with $x^i$ ($i=1,2,3$) the 
spatial coordinates on these $t=$~const. hypersurfaces, the Bianchi 
line element in comoving coordinates is 
\cite{EMMacC,Ellis:1968vb,Pontzen:2007ii}
\be
ds^2=-dt^2 + \gamma_{ij}(t) \, \mbox{e}^{\beta_i(t)} \, dx^i dx^j \quad 
\quad (i,j=1,2,3)
\ee
with $g_{00}=-1 $, $g_{0i}=0$ (that is, comoving and synchronous 
coordinates coincide), and where $ \sigma_{ij}=\sigma_{ij}(t) $. 
Equation~(\ref{quella}) then gives again that $\nabla_a X$ is parallel to 
$u_a$, $h_{ab} \dot{u}^b=0$, and $q_a=0$ (moreover, in covariant 
notation, $h_{ab} \nabla^b P= h_{ab}\nabla^b \sigma_{cd}=0$).

Let us consider explicitly Bianchi~I universes for illustration. 
Bianchi~I  models sourced by a single anisotropic fluid have line element 
\cite{EMMacC,Ellis:1968vb,Ganguly:2021pke,Pontzen:2007ii}
\be
ds^2 =- dt^2 + a^2(t) \, \mbox{e}^{ 2\beta_{(i)}(t) } \delta_{ij} dx^i 
dx^j  \quad \quad (i,j=1,2,3)
\ee
in comoving coordinates. If the single fluid sourcing the 
Bianchi~I universe is the effective Horndeski $\phi$-fluid ({\em i.e.}, in 
vacuum Horndeski cosmology), using Eq.~(\ref{quella}) we find 
again $ \dot{u}^b $ parallel to $u^b$ and $q_a=0$.  A direct 
computation gives $q_a=0$ ({\em e.g.}, 
\cite{EMMacC,Ganguly:2021pke,Pontzen:2007ii}). 
This result still holds in the 
presence of a real fluid if its 4-velocity is aligned with $u^a$.

Let us consider now Horndeski gravity in the presence of matter, which is 
usually taken to be a fluid. If this fluid is not tilted with respect to 
the Horndeski effective fluid, then its 4-velocity coincides with $u^a$ 
given by Eq.~(\ref{4-velocity}) and the previous arguments apply again. 
This is not the case, in general, if the real fluid is tilted with respect 
to the effective one.

The discussion of this section legitimates the study of FLRW and Bianchi 
cosmology in the first-order thermodynamics of scalar-tensor gravity. 
These discussions draw conclusions based on $ K{\cal T}$ 
\cite{Giardino:2022sdv}  
even though the heat flux vector~(\ref{questa}) (from which $ K{\cal 
T}$ is derived) vanishes identically in the comoving frame of the 
Horndeski effective fluid.\footnote{An observer in a frame moving with 
respect to $u^a$ would see this effective fluid tilted and would  
experience a non-vanishing energy flux $q_a '$ which is, however, purely 
convective \cite{EMMacC,Miranda:2022uyk}.} The reason why $q^a$ vanishes 
is not because $ K {\cal T}$ is zero (which would invalidate the 
discussions of Eckart's  
thermodynamics in cosmology), but because $\dot{u}^a$ is parallel to the 
trajectories of fluid particles ({\em i.e.}, to $u^a$). (It is 
possible that $ K{\cal T}$ is always zero in a certain 
specific Horndeski theory because $G_{4\phi}-XG_{3X}$ in 
Eq.~(\ref{KTHorndeski}) vanishes identically there, which makes 
this theory with non-dynamical $\phi$ a state of equilibrium alternative 
to GR \cite{Miranda:2022wkz}.)

\section{Decomposition of any symmetric tensor in the ``imperfect fluid'' 
form}
\label{sec:4}

In order to appreciate the first-order thermodynamics of scalar-tensor or 
viable Horndeski gravity 
\cite{Faraoni:2018qdr,Faraoni:2021lfc,Faraoni:2021jri,Giusti:2021sku,Faraoni:2022gry,Giardino:2022sdv,Giusti:2022tgq,Faraoni:2022doe,Faraoni:2022jyd,Faraoni:2022fxo,Miranda:2022uyk},
one should understand what is peculiar to the effective stress-energy 
tensor of these theories, once their field equations are written as 
effective Einstein equations. It is not the fact that their effective 
stress-energy tensor assumes the form of a dissipative fluid---this is 
true for {\em any} symmetric 2-index tensor. What is peculiar is the fact 
that this effective stress-energy tensor {\em satisfies the constitutive 
relations of Eckart's first-order thermodynamics}. This property is truly 
remarkable and is certainly not warranted.\footnote{See 
\cite{Miranda:2022wkz} for an attempt to generalize the analogy to a 
non-Newtonian fluid with alternative constitutive relations non-linear in 
the gradient 
of the fluid 4-velocity.} Let us discuss explicitly the dissipative fluid 
decomposition of a symmetric tensor.

Given a timelike vector field $u^a$ normalized so that $u_c u^c=-1$, 
{\em any} symmetric 2-index tensor $S_{ab}=S_{ba}$ can be decomposed in 
the imperfect fluid form
\be
S_{ab}=\rho u_a u_b +P h_{ab} + q_a u_b + q_b u_a + \pi_{ab} \,,
\ee
where $ 
h_{ab} \equiv g_{ab}+u_a u_b  $ 
and $q^a$  and $\pi^{ab}$ are purely spatial with respect to $u^a$, with 
$\pi^{ab}$ symmetric 
and trace-free. This ``imperfect fluid decomposition'' is purely 
formal since, in general, the symmetric tensor $S_{ab}$ is not a real 
or effective stress-energy tensor, and does not even have the dimensions 
of stress-energy.

In general, the constitutive relations of Eckart's first-order 
thermodynamics~(\ref{Eckart1})-(\ref{Eckart3}) are 
 not satisfied by the components of $S_{ab}$, and neither is any other 
prescribed constitutive relation. By contrast, the effective 
stress-energy tensor of  scalar-tensor gravity $T_{ab}^{(\phi)}$ satisfies 
Eckart's constitutive relations, as does that of a restricted class of 
Horndeski theories of gravity \cite{Giusti:2021sku,Miranda:2022wkz}. In 
general, given an alternative theory of gravity {\em in vacuo}, one can 
rewrite its field equations as effective Einstein equations with a 
suitable, symmetric, effective stress-energy tensor $
T_{ab}^\mathrm{(eff)}$. However:

\begin{enumerate}

\item In general, a preferred 4-velocity vector field $u^a$ is not 
defined. If it is defined, as in scalar-tensor or Horndeski gravity where 
there is a scalar field $\phi$ and $\nabla^a \phi$ singles out a preferred 
vector field, the fluid-dynamical analogy requires that 

	\begin{itemize}
	\item $\nabla^c \phi$ is timelike, $\nabla^c\phi \nabla_c \phi 
	<0$;

	\item $\nabla^c\phi $ is future-oriented, $ g_{ab} \nabla^a \left( 
\partial_t \, \right)^b  <0$.

	\end{itemize}

\item If a preferred (timelike, normalized, and future-oriented) vector 
field is not present in the alternative theory of gravity, one could 
choose one arbitrarily, which corresponds to choosing a family of physical 
observers in spacetime. Then, provided that the field equations of this 
theory can be written as effective Einstein equations, one has an 
effective symmetric $T_{ab}^\mathrm{(eff)}$ which can be decomposed in the 
form of an imperfect fluid. However, $u^c$ has no relation with the 
gravitational degrees of freedom of the theory and, in general, no 
constitutive relation is satisfied. This is intuitive: constitutive 
relations express the physical properties of a material (specifically, its 
response to mechanical and thermal stresses) and there is no physics in 
the purely geometric decomposition of a tensor into its temporal, spatial, 
and mixed components. It is remarkable that scalar-tensor gravity 
does indeed satisfy Eckart's  constitutive relations.

\end{enumerate}

\subsection{Decomposition}

Let $S_{ab} $ be any symmetric 2-index tensor in a spacetime endowed with 
a metric $g_{ab}$ and let $u^a$ be a timelike vector field. Without loss 
of generality, we can assume that $u^a$ is normalized to $u_c u^c=-1$ 
(otherwise one can always normalize it). Define the 3-metric $
h_{ab} \equiv u_a u_b + g_{ab} $
~(${h_a}^b$ is the projector onto the 3-space seen by $u^a$, {\em i.e.}, 
$h_{ab}u^a =h_{ab} u^b=0$). Then is is always possible to decompose 
$S_{ab}$ according to 
\be
S_{ab} =  \rho u_a u_b + P h_{ab} + q_a u_b + q_b u_a +\pi_{ab} \,,
\label{decomposition}
\ee
where 
\be
q_a u^a =0 \,, \quad \quad \pi_{ab} u^a = \pi_{ab} u^b =0 \,, \quad \quad 
{\pi^a}_a=0 \,.
\ee
The quantities appearing in this decomposition are
\begin{eqnarray}
\rho & = & S_{ab} u^a u^b \,,\label{density}\\
&&\nonumber\\
P &=& \frac{1}{3} \, h^{ab} S_{ab} \,,\label{pressure}\\
&&\nonumber\\
q^a  &=& - {h^a}^c S_{cd} u^d \,,\label{heatfluxdensity}\\
&&\nonumber\\
\pi_{ab}  &=& \left( {h_a}^c \, {h_b}^d -\frac{1}{3} \,  h_{ab} \, h^{cd} 
\right) S_{cd} \,.\label{anisotropicstresses}
\end{eqnarray}
They are just  the projections of $S_{ab}$ onto the time 
direction (projected twice for $\rho$), onto the 3-space (projected twice 
for the isotropic 
and anisotropic stresses $P h_{ab}$ and 
$\pi_{ab}$), and projected once onto the 3-space/once onto the time 
direction (for 
$q^a$). In this sense, the decomposition is rather obvious 
(it is 
mentioned, {\em e.g.},  in~\cite{EMMacC,Apostolopoulos:2016zbfw}, but 
seems to have been missed by many authors discussing various 
scalar-tensor theories over the years).

\noindent {\em Proof.} By definition, $q^a$ and $\pi^{ab}$ are purely 
spatial since they are a projection and a double projection onto the 
3-space seen by $u^a$:
$$
q_a u^a \equiv - {h_a}^c \left( S_{cd} u^d \right) u^a =0 
$$
and
\begin{eqnarray*}
\pi_{ab} u^a & \equiv & \left( {h_a}^c \, {h_b}^d -\frac{1}{3} \, h_{ab} 
\, h^{cd} \right) S_{cd} u^a \nonumber\\
&&\nonumber\\
&=& -\frac{1}{3} \, \left( h^{cd} 
S_{cd} 
\right) h_{ab} u^a=0 \,,\\
&&\nonumber\\
\pi_{ab} u^b & \equiv & \left( {h_a}^c \, {h_b}^d -\frac{1}{3} \, h_{ab} 
\, h^{cd} \right) S_{cd} u^b \nonumber\\
&&\nonumber\\
&=& -\frac{1}{3} \, \left( h^{cd} 
S_{cd}  \right) h_{ab} u^b=0 \,,
\end{eqnarray*}
and
$$
{\pi^a}_a = \left( {h_a}^c \, h^{ad} -\frac{h}{3} \, h^{cd} \right) S_{cd} 
=\left( h^{cd} -\frac{3}{3} \, h^{cd} \right) S_{cd}=0 \,.
$$
It is easy to show that, using the 
quantities~(\ref{density})-(\ref{anisotropicstresses}), the 
right-hand side of Eq.~(\ref{decomposition}) reproduces the given tensor 
$S_{ab}$. In fact,
\begin{eqnarray*}
&\, & \rho u_a u_b  + P h_{ab} + q_a u_b + q_b u_a +\pi_{ab} \equiv   
\left( S_{cd} u^c u^d \right) u_a u_b \nonumber\\
&&\nonumber\\
&\, & + \left( \frac{ S_{cd} 
h^{cd}}{3} 
\right) h_{ab}  - \left({h_a}^c S_{cd} u^d \right) u_b 
 \nonumber\\
&&\nonumber\\
&\, & - \left({h_b}^c S_{cd} u^d \right) u_a 
+ \left( {h_a}^c \, {h_b}^d -\frac{ h^{cd} S_{cd}}{3} \right) S_{cd} 
\nonumber\\
&&\nonumber\\
& = &   
\left( S_{cd} u^c u^d \right) u_a u_b + \left( \frac{ S_{cd} h^{cd}}{3} 
\right) h_{ab} 
- \left( {\delta_a}^c + u_a u^c \right)  \left( S_{cd} u^d \right) u_b  
\nonumber\\
&&\nonumber\\ 
&\, & - \left( {\delta_b}^c + u_b u^c \right)  \left( S_{cd} u^d \right) 
u_a 
+ \left( {\delta_a}^c + u_a u^c \right)\left( {\delta_b}^d + u_b u^d 
\right)  S_{cd} \nonumber\\
&&\nonumber\\
&\, & -\frac{h_{ab}}{3} \left(
g^{cd} + u^c u^d \right) S_{cd}  \nonumber\\
&&\nonumber\\
& = &  \left( S_{cd} u^c u^d \right) u_a u_b 
+ \left( \frac{ S_{cd} h^{cd}}{3} \right) h_{ab} 
-\left( S_{ad} u^d \right) u_b \nonumber\\
&&\nonumber\\
&\, & -\left( S_{cd} u^c u^d \right) u_a u_b 
-\left( S_{bd} u^d \right) u_a \nonumber\\
&&\nonumber\\ 
&\, & 
-\left( S_{cd} u^c u^d \right) u_a u_b 
+S_{ab}
+\left( S_{ad} u^d \right) u_b
+ \left( S_{cb} u^c \right) u_a \nonumber\\
&&\nonumber\\
&\, & + \left( S_{cd} u^c u^d \right) u_a u_b 
-\frac{S}{3} \, h_{ab}
- \frac{S_{cd} u^c u^d}{3} \, h_{ab} \nonumber\\
&&\nonumber\\
& = &  S_{ab} +\frac{h_{ab}}{3} \left[ S_{cd}\left( g^{cd} +u^c 
u^d \right) -S -S_{cd} u^c u^d \right] \nonumber\\
&&\nonumber\\
& = & S_{ab} \,.
\end{eqnarray*}

\subsection{Effective constitutive relations}

Apart from the fact that they do not have the dimensions of fluid 
quantities, in general the quantities appearing in the effective 
dissipative fluid decomposition do not satisfy effective constitutive 
relations. For example, the first of Eckart's  constitutive 
relations~(\ref{Eckart1}) corresponds to
\be
S_{0i} = - K h_{ij} \left( \partial^j {\cal T} +{\cal T} \dot{u}^j \right)
\ee
and one cannot see how functions $K$ and ${\cal T}$ could exist to 
satisfy this relation between $S_{ab}$ and the acceleration $\dot{u}^a$. 
Similarly, Eq.~(\ref{Eckart2}) corresponds to 
\be
\left( {h_i}^c \, {h_j}^d - \frac{h_{ij}}{3} \, h^{cd} \right) S_{cd} = 
-2\eta \left( \nabla_{(i} u_{j)} -\frac{\nabla_c u^c}{3} \, h_{ij} \right) 
\,,
\ee
which is impossible to satisfy in general if $S_{ab}$ does not have a 
special relation with $u^a$ and $\dot{u}^a$ as it happens instead in 
scalar-tensor gravity, where $u^c$ is the (normalized) gradient of the 
gravitational scalar degree of freedom $\phi$ and $S_{ab}=T_{ab}^{(\phi)}$ 
is built out of $\phi$ and its derivatives.

\subsection{Examples}

As the first example of the imperfect fluid decomposition of  asymmetric 
tensor, consider the metric itself, $S_{ab}=g_{ab}$ (``imperfect fluid'' 
decomposition is just a name here since the dimensions of $g_{ab}$ are not 
those of a stress-energy tensor). The formal imperfect fluid quantities 
are
\begin{eqnarray}
\rho^\mathrm{(g)} & = & g_{ab} u^a u^b =-1 \,,\\
&&\nonumber\\
P^\mathrm{(g)} & = & \frac{1}{3} \, h^{ab} g_{ab} = 1 \,,\\
&&\nonumber\\
q_a^\mathrm{(g)} & = & -{h_a}^c \,  g_{cd} u^d = 
 -{h_a}^c \, u_c =0 \,,\\
&&\nonumber\\
\pi_{ab}^\mathrm{(g)} & = &
\left(  {h_a}^c \,  {h_b}^d  -\frac{1}{3} \, h_{ab}\, h^{cd}\right) h_{cd} 
=  h_{ad}\, {h_b}^d -\frac{3}{3} \, h_{ab} =0 \,.\nonumber\\
&&
\end{eqnarray}
The corresponding imperfect ``fluid'' reduces to a perfect one with 
equation of state $P_\mathrm{(g)}= - \rho_\mathrm{(g)}$. Indeed, 
the cosmological constant term $\Lambda g_{ab}$ in the Einstein equations
\be
R_{ab} -\frac{1}{2} \, g_{ab} R +\Lambda g_{ab} =8\pi G 
\, T_{ab}^\mathrm{(matter)}
\ee
can be seen as an effective fluid with stress-energy tensor 
$T_{ab}^{(\Lambda)} =-\frac{ \Lambda}{8\pi G} \, g_{ab}$ and with the 
properties above. In addition, the constants $\Lambda$ and $G$ in 
$S_{ab}=-\frac{\Lambda}{8\pi G} \, g_{ab}$ give this tensor the correct 
dimensions for a stress-energy tensor.

As a second example consider the Ricci tensor, $S_{ab} = R_{ab}$. The 
effective 
dissipative fluid quantities are related to the components of $R_{ab}$ 
in the frame of the observers with 4-velocity  $u^a$:
\begin{eqnarray}
\rho^\mathrm{(Ricci)} &=& R_{ab} u^a u^b =R_{00} \,,\\
&&\nonumber\\
P^\mathrm{(Ricci)} &=& \frac{1}{3} \, h^{ab} R_{ab}  \,,\\
&&\nonumber\\
q_i^\mathrm{(Ricci)} &=& - {h_i}^c  R_{cd} u^d= -  R_{i0} \,,\\
&&\nonumber\\
\pi_{ij}^\mathrm{(Ricci)} &=& \left( {h_i}^c \, {h_j}^d -  
\frac{h_{ij}}{3} \, h^{cd} \right) R_{cd} = R_{ij}- \frac{ 
h^{cd}R_{cd}}{3} \, h_{ij}   \,, \nonumber\\
&&
\end{eqnarray}
where $i,j=1,2,3$. 

Finally, any purely spatial tensor (such as the extrinsic curvature 
$K_{ij}$, the shear tensor $\sigma_{ij}$, or the 3-metric $h_{ij}$ itself) 
will have vanishing effective $\rho$ and $q_a$ and 
non-vanishing effective ``stresses'' (including $P$ and $\sigma_{ab}$).

\section{Conclusions}
\label{sec:5}

The correct Eckart generalization of the Fourier law is important for the 
study of the first-order thermodynamics of scalar-tensor gravity in 
cosmology, which is now made completely legitimate by our considerations 
of Sec.~\ref{sec:2} and Sec.~\ref{sec:3}.  The discussion has been 
extended to include spatially homogeneous and anisotropic Bianchi 
universes, not discussed before. The analysis of specific Bianchi models 
with regard to the general thermodynamical ideas advanced in previous 
publications involves phase space analyses and much detail and will be 
pursued elsewhere.

A key point of the first-order thermodynamics of scalar-tensor gravity is 
often misunderstood and has not been spelled out thus far. Writing the 
field equations of scalar-tensor gravity as effective Einstein equations 
produces an effective stress-energy tensor $T_{ab}^{(\phi)}$ as a source. 
The latter has the form~(\ref{imperfectfluid}) of an imperfect fluid 
energy-momentum tensor, but this fact contains no physics: {\em any} 
symmetric two-index tensor admits this decomposition, which is purely 
mathematical. It is the almost miracolous fact that the effective 
$\phi$-fluid quantities thus derived satisfy Eckart's constitutive 
relations (which, in non-relativistic physics, characterize a Newtonian 
fluid) that make the first-order thermodynamics work.

\begin{acknowledgements}

This work is supported by the Natural Sciences \& Engineering Research 
Council of Canada (grant 2016-03803 to V.F.), Fondation Arbour, and a 
Bishop's University Graduate Entrance Scholarship (J.H.).

\end{acknowledgements}

\bigskip
\appendix
\section{Force parallel to a worldline}
\label{appendix:A}
\renewcommand{\theequation}{A.\arabic{equation}}

When a 4-force parallel to the wordline of a particle ({\em i.e.}, to its 
4-tangent) is present, the equation of motion of this particle coincides 
with the non-affinely parametrized geodesic equation. Let $\tau$ be the 
{\em proper time} along this worldline (not an affine parameter) and let 
\be 
u^c \equiv \frac{dx^c}{d\tau} \,, \quad \quad a^c \equiv \frac{d^2 
x^c}{d\tau^2} = \frac{du^c}{d\tau} 
\ee 
according to the standard 
definitions of 4-velocity and 4-acce\-leration. In cosmology, the comoving 
time $t$ is the proper time of comoving observers but is not an affine 
parameter along their worldlines unless the cosmic fluid is a dust or a 
cosmological constant term because 
these observers are accelerated by a pressure gradient pointing in the 
direction of comoving time. Of course, one can always introduce an affine 
parameter 
along these fluid worldlines, but this is not convenient since one wants 
instead to use formulas written in comoving 
coordinates, associated with the physical observers seeing the cosmic 
microwave background homogeneous and isotropic around them (on average).

Let $s$ be an affine parameter along the cosmic fluid worldlines. We have
\begin{eqnarray}
u^c & \equiv &  \frac{dx^c}{d\tau} = \frac{dx^c}{ds} \, \frac{ds}{d\tau} 
\,,\\
&&\nonumber\\
a^c & \equiv &  \frac{d^2x^c}{d\tau^2} = \frac{du^c}{d\tau} =
\frac{d}{d\tau} \left( \frac{dx^c}{ds} \, \frac{ds}{d\tau} \right) \\
&&\nonumber\\
&=& \frac{d^2 x^c}{d\tau \,ds} \, \frac{ds}{d\tau} + 
\frac{d x^c}{ds} \, \frac{d^2s}{d\tau^2}\nonumber\\
&&\nonumber\\
&=& \left[ \frac{d}{d\tau} \left( \frac{dx^c}{ds} \right) \right] 
\frac{ds}{d\tau} + \frac{dx^c}{d\tau} \, \frac{d\tau}{ds} \, 
\frac{d^2s}{d\tau^2} \nonumber\\
&&\nonumber\\
&=& \frac{ds}{d\tau} \left[ \frac{d}{ds} \left( \frac{dx^c}{ds}\right) 
\right] \frac{ds}{d\tau} + u^c \, \frac{d\tau}{ds}\, \frac{d^2 s}{d\tau^2}   
\,,
\end{eqnarray}
and
\be
a^c = \frac{d^2 x^c}{ds^2} \left( \frac{ds}{d\tau} \right)^2 + u^c \, 
\frac{d\tau}{ds} \, \frac{d^2 s}{d\tau^2} \,, \label{zac1}
\ee
or
\be
\frac{d^2 x^c}{ds^2} = a^c \left( \frac{d\tau}{ds} \right)^2 -u^c \left( 
\frac{d\tau}{ds} \right)^3 \frac{d^2s}{d\tau^2} \,. 
\ee

Now, since $s$ is an affine parameter along the wordline,
\be
\frac{d^2 x^c}{ds^2}  + \Gamma^c_{ab} \, \frac{dx^a}{ds} \, 
\frac{dx^b}{ds} =0 \,,
\ee

\noindent or
\begin{eqnarray}
a^c \left( \frac{d\tau}{ds} \right)^2 
- u^c \left( \frac{d\tau}{ds}\right)^3 
\frac{d^2s}{d\tau^2} +\Gamma^c_{ab} \, \frac{dx^a}{d\tau} \, 
\frac{dx^b}{d\tau} \left( \frac{d\tau}{ds} \right)^2 =0 \nonumber\\
\end{eqnarray}
and
\be
a^c +\Gamma^c_{ab} \, \frac{dx^a}{d\tau} \, \frac{dx^b}{d\tau} =
u^c \, \frac{d\tau}{ds} \, \frac{d^2s}{d\tau^2} \,.\label{eq:azzo}
\ee

We also have 
\begin{eqnarray}
\frac{d^2\tau}{ds^2} &=& \frac{d}{ds} \left( \frac{1}{ds/d\tau} \right) = 
\frac{d\tau}{ds} \frac{d}{d\tau} \left(  \frac{1}{ds/d\tau} \right) 
\nonumber\\
&&\nonumber\\
&=& 
-\frac{d\tau}{ds} \, \frac{ d^2s/d\tau^2}{\left( ds/d\tau  \right)^2} =
- \left( \frac{d\tau}{ds}  \right)^3
\frac{ d^2s}{d\tau^2} \,,
\end{eqnarray}
then Eq.~(\ref{eq:azzo}) can be written also as
\be
a^c +\Gamma^c_{ab} \, \frac{dx^a}{d\tau} \, \frac{dx^b}{d\tau} =
- u^c \left(  \frac{d\tau}{ds} \right)^{-2} \frac{d^2 \tau}{d s^2} \,.
\ee

The orthogonality of the 4-acceleration $a^c$ to the 
4-velocity engraved in the mind of relativists, $a^c u_c=0$, follows 
from differentiating the normalization $u^c u_c=-1$, but $dx^c/ds$ is not 
normalized and $g_{ab} \, \frac{d^2x^a}{ds^2} \, \frac{dx^b}{ds} \neq 0$. 
In fact, 
\begin{eqnarray}
g_{ab} \, \frac{d^2x^a}{ds^2}  \, \frac{dx^b}{ds} &=& g_{ab} 
\left( a^a  -u^a \, \frac{d\tau}{ds} \, \frac{d^2s}{d\tau^2} \right) 
\frac{ 1}{ 
\left( ds/d\tau \right)^2}  \frac{ dx^b}{d\tau} \, \frac{d\tau}{ds} 
\nonumber\\
&=& - g_{ab} u^a u^b \, \frac{d^2s}{d\tau^2} \left( \frac{d\tau}{ds} 
\right)^3 \frac{d\tau}{ds} \nonumber\\
&&\nonumber\\
& = & \left( \frac{d\tau}{ds} \right)^4 
\frac{d^2s}{d\tau^2} \,,
\end{eqnarray}
which is different from zero unless $\tau$ is already an affine parameter. 
Likewise, we have
\begin{eqnarray}
&& g_{ab} \, \frac{dx^a}{ds} \, \frac{dx^b}{ds}  =
g_{ab} \, \frac{dx^a}{d\tau} \, \frac{dx^b}{d\tau} \, \left( 
\frac{d\tau}{ds} \right)^2 =-\left( \frac{d\tau}{ds} \right)^2 \neq -1 
\,.\nonumber\\
&&
\end{eqnarray}



\end{document}